\documentclass[aps,prd]{revtex4}
\usepackage{graphicx}
\usepackage{amssymb}
\def\abstract#1{\vskip 7mm 
        \begin{center}{\large Abstract}\par \smallskip
                \begin{minipage}[c]{12cm}
                        \small #1
                \end{minipage}
        \end{center}
}
\def\title#1{\begin{center}{\Large\bf #1}\end{center}}
\def\author#1{\vskip 5mm \begin{center}{#1}\end{center}}
\def\address#1{\begin{center}{\it #1}\end{center}}
\def\lsim{\mathrel{\rlap{\lower4pt\hbox{\hskip1pt$\sim$}}
    \raise1pt\hbox{$<$}}}                
\def\gsim{\mathrel{\rlap{\lower4pt\hbox{\hskip1pt$\sim$}}
    \raise1pt\hbox{$>$}}}                

\makeatletter
\@ifundefined{lesssim}{}{}
\@ifundefined{gtrsim}{}{}
\def\vereq#1#2{\lower3pt\vbox{\baselineskip1.5pt \lineskip1.5pt
\ialign{$\m@th#1\hfill##\hfil$\crcr#2\crcr\sim\crcr}}}
\makeatother

\begin{document}

\title{%
Spectroscopy of Cosmic topology
}
\author{%
{\bf Tarun Souradeep} 
}
\address{%
Inter-University Centre for   Astronomy and Astrophysics,\\ 
Post Bag 4, Ganeshkhind, Pune 411007,  India\\
{E-mail:tarun@iucaa.ernet.in}}

\abstract{ Einstein's theory of gravitation that governs the geometry
of space-time, coupled with spectacular advance in cosmological
observations, promises to deliver a `standard model' of cosmology in
the near future.  However, local geometry of space constrains, but
does not dictate the topology of the cosmos. hence, Cosmic topology
has remained an enigmatic aspect of the `standard model' of
cosmology. Recent advance in the quantity and quality of observations
has brought this issue within the realm of observational query.  The
breakdown of statistical homogeneity and isotropy of cosmic
perturbations is a generic consequence of non trivial cosmic topology
arising from to the imposed `crystallographic' periodicity on the
eigenstates of the Laplacian. The sky maps of Cosmic Microwave
Background (CMB) anisotropy and polarization most promising
observations that would carry signatures of a violation of statistical
isotropy and homogeneity.  Hence, a {\em measurable} spectroscopy of
cosmic topology is made possible using the Bipolar power spectrum
(BiPS) of the temperature and polarization that quantifies violation
of statistical isotropy.  }


\footnotetext[1]{Invited contribution to a special issue of IJP in
 memory of Prof. A.K. Raychaudhuri. }

\section{Introduction}

I feel honored to be invited to contribute to this volume honoring the
memory of {\bf Professor Amal Kumar Raychaudhuri} (fondly known as
AKR) -- a great scientist and teacher. The {\em Raychaudhuri
equations} describing the evolution of anisotropic universe models are
the footprints of homegrown Indian science in the field of
cosmology. Though the background universe is observationally
consistent with homogeneous and isotropic Friedmann models, the
Raychaudhuri equations appears in the evolution of inhomogeneities
that led to the formation of large scale structures in the
universe. It is fair to say much of the recent progress in cosmology
has come from the interplay between refinement of the theories of
structure formation and the improvement of the observations. Hence,
the Raychaudhuri equations have remained as relevant and ingrained in
contemporary cosmology as when first put forward by AKR. This article
describing our ongoing research determine the topology of the universe
from the exquisite measurements of anisotropy in the Cosmic Microwave
background is a humble tribute to the doyen of Indian science.

\begin{figure}[h]
  \begin{center}
    \includegraphics[scale=0.5]{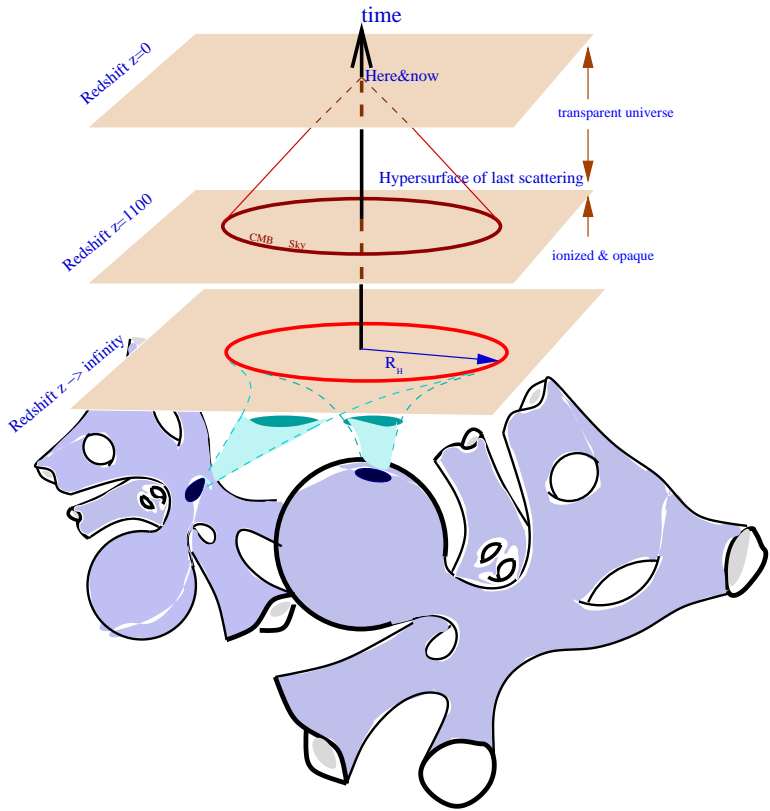}
  \end{center}
  \caption{\footnotesize A cartoon depicting a prevalent view within the
  inflationary paradigm. The observable universe corresponds to a
  small patch of a very complicated manifold that has been blown to
  cosmological scales during an inflationary epoch. Ultra-large scale
  structure could be observable if the the size of this patch is not
  much smaller that the scales of inhomogeneity and non-trivial
  topology.}
  \label{ulss_infl}
\end{figure}

The realization that a universe with the same local geometry has many
different choices of global topology has been a theoretical curiosity
as old as modern cosmology. De Sitter was quick to point out that the
first modern model of the cosmos, Einstein's closed ($S^3$, spherical
geometry and static) universe model, could equally well correspond to
the multiply connected `Elliptical' universe where antipodal point of
$S^3$ are topologically identified ($S^3/Z_2$). Fig.~\ref{ulss_infl}
depicts the prevalent modern view within the concept of inflation,
that this relatively smooth `Hubble volume' that we observe is perhaps
a tiny patch of an extremely inhomogeneous and complex spatial
manifold.  The complexity could involve non-trivial topology (multiple
connectivity) on these ultra-large scales. Given the observational
support for a homogeneous Hubble volume around us, the diverse
possibility of global structure reduces to the tractable study limited
to spaces of uniform curvature (locally homogeneous and isotropic FLRW
models) but with non-trivial cosmic topology. For example, in
Fig.~\ref{ulss_infl}, observers in the `handle' regions would perceive
an open (hyperbolic geometry ) universe, and those in the `bulb'
region observe a closed (spherical geometry) universe.  Although, in a
generic manifold, sectors with exact Euclidean geometry are rare, an
epoch of inflation in the early universe inflate any region to a nearly
flat, Euclidean geometry.

The motivation for non-trivial topology and the quest to determine
size and the shape of our universe has a rich and diverse history in
modern
cosmology~\cite{ell71,sok_shv74,got80,lac_lum95,stark98}. Spatially
compact universe models have both deep theoretical and philosophical
appeal -- eg., the quantum creation of finite sized universe from
vacuum is a theoretical motivation \cite{linde}, and a philosophical
abhorrence of any `infinity' in nature would argue against an infinite
sized universe~\cite{lev02}. A compact universe, with the exception of
the three sphere, $S^3$ implies a multiply connected universe (i.e.,
nontrivial cosmic topology).

The photons of the cosmic microwave background(CMB) propagate freely
over distances comparable to the cosmic horizon. The CMB anisotropy
and its polarization are the most promising observational probes of
the global spatial structure of the universe on length scales near to
and even somewhat beyond the `horizon' scale ($\sim c H_0^{-1}$).
{\em A generic consequence ~\footnote{Global isotropy of space is
violated in all multi-connected models (except, the `Elliptical' space
$S^3/Z_2$~\cite{sour00}).} of cosmic topology is the breaking of
statistical isotropy in characteristic patterns determined by the
photon geodesic structure of the manifold}~\cite{bps98}.  The
increasingly exquisite measurements of CMB anisotropy have brought
cosmic topology from the realm of theoretical possibility to within
the grasp of observations and has received considerable attention over
the past few years~\cite{staro,stark98,levin98, bps98,
bps00a,bps00b,angelwmap,coles-graca}.  CMB polarization not only
augments the CMB temperature anisotropy observations but is expected
to allow a more incisive study of cosmic topology since it arises only
at the surface of last scattering (SLS).  The recent full sky
measurements of CMB polarization maps by WMAP has opened the door to
this new arena~\cite{pag06}.

Any multiply-connected space is equivalent to a simply connected space
(universal cover) that is tiled by Dirichlet domains (DD) under a free
acting subgroup, $\Gamma$ of the isotropy group, $G^u$ of universal
cover. {\em Hence, even the most complicated cosmic topology reduces
to a study of standard FLRW models with periodic boundary conditions
on the appropriate DD.} A simple example being the 3-torus $T^3$, that
is equivalent to the study of a universe in a box with periodic
boundary conditions -- a routine approximation in numerical
simulations of large scale structure in the universe. More formally,
the $T^3$ is obtained by tiling 3-D Euclidean space under the isometry
subgroup of discrete translations in three directions.  In cosmology,
the Dirichlet domain constructed around the observer represents the
universe as `seen' by the observer.  The SI breakdown is apparent in
the principal axes present in the shape of the DD constructed with the
observer located at the base-point~\cite{us_prl}.  Equivalently, the
fields defining perturbations are built of the subset of eigenstates
of Laplacian that invariant under $\Gamma$.  The correlations of CMB
fluctuations in the sky would have patterns that are no longer
invariant under rotations.

\begin{figure}[h]
\centering
\includegraphics[scale=0.38,
angle=0]{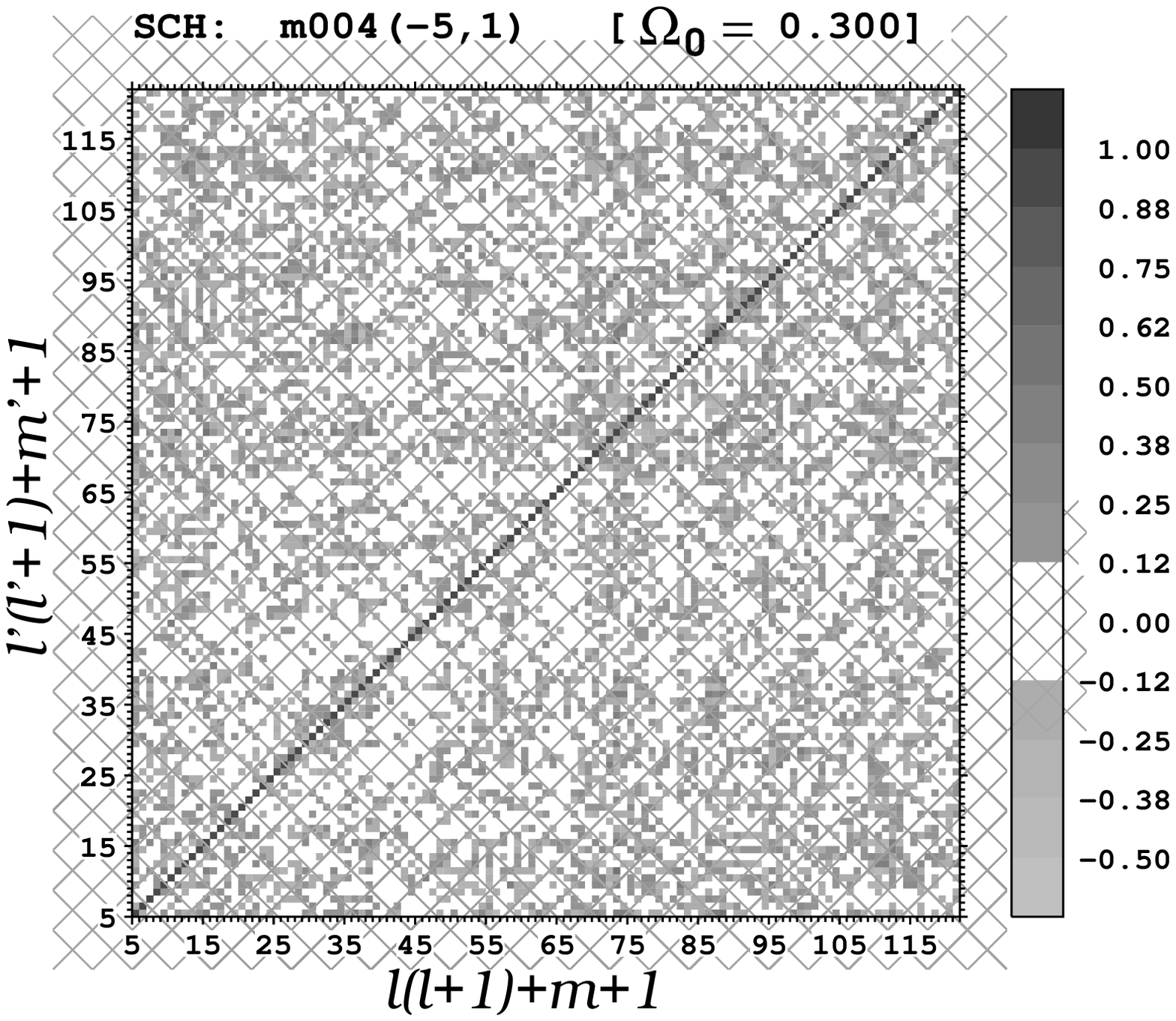}
\includegraphics[scale=0.38,
angle=0]{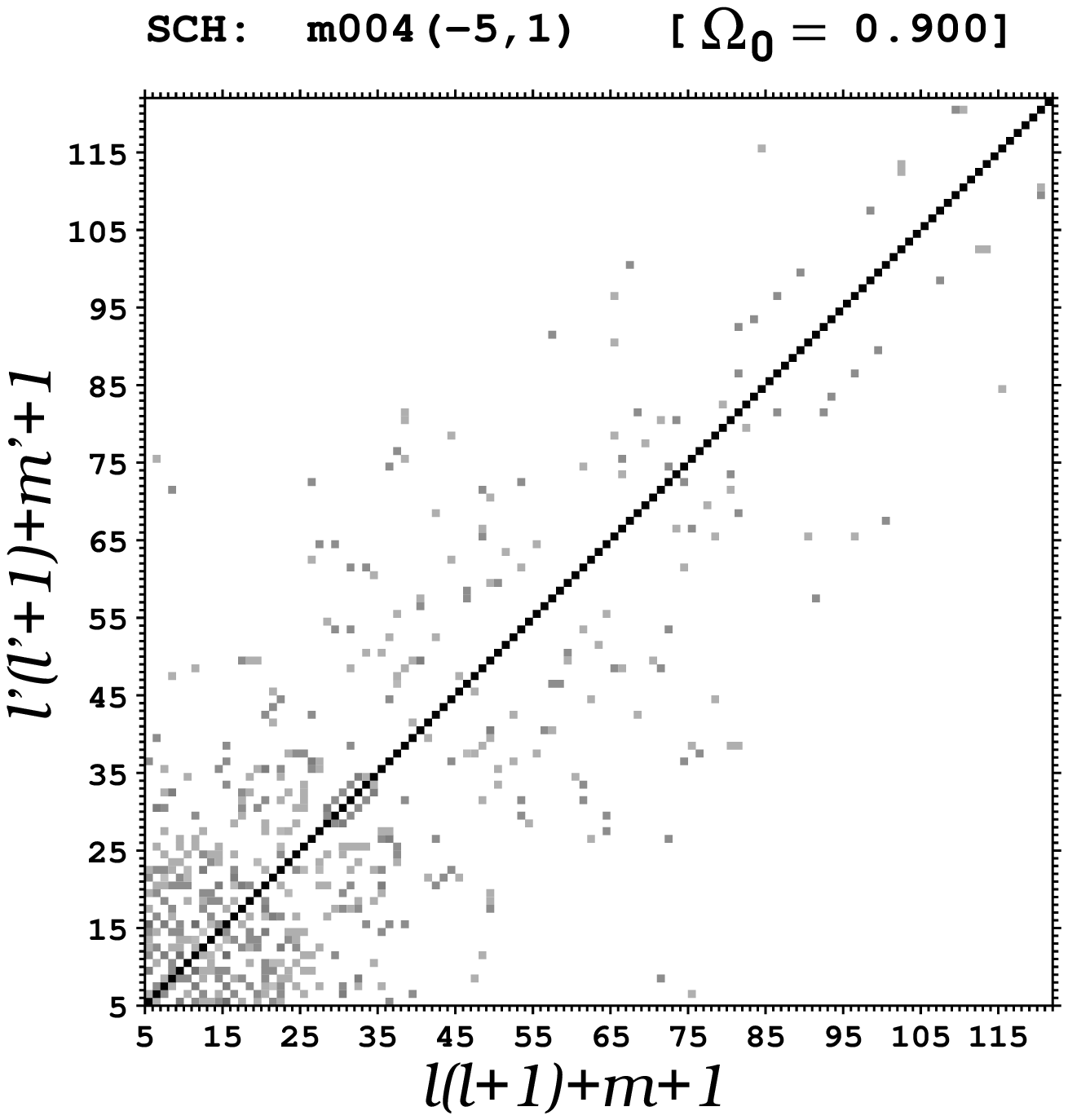}
\caption{\footnotesize The figure taken from~\protect\cite{bps00b}
illustrates the non-diagonal nature of the expectation values of
$a_{\ell m}$ pair products when the CMB anisotropy violates SI in two
model compact universe. The radical violation in the model on the left
corresponds to a small compact universe where CMB photons have
traversed across multiple times. The model on the left with mild
violation of SI corresponds to a universe of size comparable to the
observable horizon. For more details, see~\protect\cite{bps00b}}
\label{fig:cross_alm}
\end{figure}

In a cosmological model with trivial topology, the CMB anisotropy
signal is expected to be statistically isotropic, i.e., statistical
expectation values of the temperature fluctuations $\Delta T(\hat q)$
are preserved under rotations of the sky. In particular, the angular
correlation function $C(\hat{q},\, \hat{q}^\prime)\equiv\langle\Delta
T(\hat q)\Delta T(\hat q^\prime)\rangle$ is rotationally invariant for
Gaussian fields, i.e., $C(\hat{q},\, \hat{q}^\prime)\equiv
C(\hat{q}\cdot\hat{q}^\prime)$.  In spherical harmonic space, where
$\Delta T(\hat q)= \sum_{lm}a_{lm} Y_{lm}(\hat q)$, the condition of
{\em statistical isotropy} (SI) translates to a diagonal $\langle
a_{lm} a^*_{l^\prime m^\prime}\rangle=C_{l}
\delta_{ll^\prime}\delta_{mm^\prime}$ where $C_l$, the widely used
angular power spectrum of CMB anisotropy.  The correlation patterns in
the CMB anisotropy that lead to violation of SI implies that imply
$\langle\hat a_{l m} \hat a^*_{l m} \rangle$ has off-diagonal
elements. Figure~\ref{fig:cross_alm} taken from ~\cite{bps00b} shows
the off-diagonal elements in the CMB correlation for two compact
universe models.

The observed CMB sky $\widetilde{\Delta T}(\hat n)$ is a single
realization of the underlying correlation, hence the detection of SI
violation or correlation patterns pose a great observational
challenge~\footnote{The same line of reasoning holds for CMB
polarization when expressed in terms of Gaussian scalar
$\widetilde{E}(\hat n)$ and pseudo-scalar $\widetilde{B}(\hat n)$
fields with corresponding two point correlation functions. For
brevity, we explicitly discuss only the temperature anisotropy.}.  For
statistically isotropic CMB sky, the correlation function
\begin{equation}
C(\hat{n}_1,\hat{n}_2)\equiv C(\hat{n}_1\cdot\hat{n}_2) =
\frac{1}{8\pi^2}\int d{\mathcal R} \,\,C({\mathcal R}\hat{n}_1,\,
{\mathcal R}\hat{n}_2),
\label{avg_cth}
\end{equation}
where ${\mathcal R}\hat{n}$ denotes the direction obtained under the
action of a rotation ${\mathcal R}$ on $\hat{n}$, and $d{\mathcal R}$
is a volume element of the three-dimensional rotation group.  The
invariance of the underlying statistics under rotation allows the
estimation of $C(\hat{n}_1\cdot\hat{n}_2)$ using the average of the
temperature product $\widetilde{\Delta T}(\hat n) \widetilde{\Delta
T}(\hat n')$ between all pairs of pixels with the angular separation
$\theta$.  In the absence of statistical isotropy,
$C(\hat{n},\hat{n}')$ is estimated by a single product
$\widetilde{\Delta T}(\hat n)\widetilde{\Delta T}(\hat n')$ and,
hence, is poorly determined from a single realization.

Although it is not possible to estimate each element of the full
correlation function $C(\hat{n},\hat{n}')$, some measures of
statistical anisotropy of the CMB map can be estimated through
suitably weighted angular averages of $\widetilde{\Delta T}(\hat
n)\widetilde{\Delta T}(\hat n')$. The angular averaging procedure
should be such that the measure involves averaging over sufficient
number of independent `measurements', but should ensure that the
averaging does not erase all the signature of statistical anisotropy.
Recently, we proposed the Bipolar Power spectrum (BiPS) $\kappa_\ell$
($\ell=1,2,3, \ldots$) of the CMB map as a statistical tool of
detecting and measuring departure from SI~\cite{us_apjl}. The BiPS is
formally defined as
 \begin{equation}\label{kl}
 \kappa^{\ell}\,=\, (2l+1)^2 \int d\Omega_{n_1}\int d\Omega_{n_2} \,
 [\frac{1}{8\pi^2}\int d{\mathcal R} \chi^{\ell}({\mathcal R})\, C({\mathcal R}\hat{n}_1,\, {\mathcal R}\hat{n}_2)]^2.
 \end{equation}
 In the above expression, $ C({\mathcal R}\hat{n}_1,\, {\mathcal
 R}\hat{n}_2)$ is the two point correlation at ${\mathcal
 R}\hat{n}_1\,$ and $ {\mathcal R}\hat{n}_2$ which are the coordinates
 of the two pixels $\hat{n}_1\,$ and $\hat{n}_2$ after rotating the
 coordinate system by element ${\mathcal R}$ of the rotation group.

 $\chi^{\ell}({\mathcal R})$ is the trace of the finite rotation
 matrix in the $\ell M$-representation
 \begin{equation}\label{chil} 
 \chi^{\ell}({\mathcal R})\,=\,\sum_{M=-\ell}^{\ell}
 D_{MM}^{\ell}({\mathcal R}),
 \end{equation}
which is called the {\it characteristic function}, or the character of
the irreducible representation of rank $\ell$. It is invariant under
rotations of the coordinate systems.  in eq.(\ref{kl}) $d{\mathcal R}$
is the volume element of the three-dimensional rotation group.  For a
statistically isotropic model $ C(\hat{n}_1,\, \hat{n}_2)$ is
invariant under rotation, and therefore $ C({\mathcal R}\hat{n}_1,\,
{\mathcal R}\hat{n}_2)\,=\,C(\hat{n}_1,\, \hat{n}_2)$ and the
orthonormality of $\chi^{\ell}(\omega)$, we will recover the condition
for SI,
 \begin{equation} 
 \kappa^{\ell} \, = \, \kappa^0 \delta_{\ell 0}.
 \end{equation}

The Bipolar power spectrum gets it name from its interpretation in the
harmonic space.  The two point correlation of CMB anisotropies,
$C(\hat{n}_1,\, \hat{n}_2)$, is a two point function on $S^2 \times
S^2$, and hence can be expanded as
\begin{equation}\label{bipolar}
C(\hat{n}_1,\, \hat{n}_2)\, =\, \sum_{l_1,l_2,L,M} A_{l_1l_2}^{\ell M}
\,\,\{Y_{l_1}(\hat{n}_1) \otimes Y_{l_2}(\hat{n}_2)\}_{\ell M},
\end{equation}
where $A_{l_1l_2}^{\ell M}$ are coefficients of the expansion (here
after BipoSH coefficients) and $\{Y_{l_1}(\hat{n}_1) \otimes
Y_{l_2}(\hat{n}_2)\}_{\ell M}$ are the Bipolar spherical harmonics
which transform as a spherical harmonic with $\ell,\, M$ with respect
to rotations \cite{Var} given by
\begin{equation}
\{Y_{l_1}(\hat{n}_1) \otimes Y_{l_2}(\hat{n}_2)\}_{\ell M} \,=\,
\sum_{m_1m_2} {\mathcal C}_{l_1m_1l_2m_2}^{\ell M} Y_{l_1 m_1}(\hat{n}_2)Y_{l_2 m_2}(\hat{n}_2),
\end{equation}
in which ${\mathcal C}_{l_1m_1l_2m_2}^{\ell M}$ are Clebsch-Gordan
coefficients. We can inverse-transform $C(\hat{n}_1,\, \hat{n}_2)$ to
get the $A_{l_1l_2}^{\ell M}$ by multiplying both sides of
eq.(\ref{bipolar}) by $\{Y_{l'_1}(\hat{n}_1) \otimes
Y_{l'_2}(\hat{n}_2)\}_{\ell'M'}^*$ and integrating over all angles,
then the orthonormality of bipolar harmonics implies that
\begin{equation}\label{alml1l2}
A_{l_1l_2}^{\ell M} \,=\,\int d\Omega_{\hat{n}_1}\int d\Omega_{\hat{n}_2} \,
C(\hat{n}_1,\, \hat{n}_2)\, \{Y_{l_1}(\hat{n}_1) \otimes Y_{l_2}(\hat{n}_2)\}_{\ell M}^*. 
\end{equation}
The above expression and the fact that $C(\hat{n}_1,\, \hat{n}_2)$ is
symmetric under the exchange of $\hat{n}_1$ and $\hat{n}_2$ lead to
the following symmetries of $A_{l_1l_2}^{\ell M}$
\begin{eqnarray}  \label{sym}
A_{l_2l_1}^{\ell M}\,&=&\,(-1)^{(l_1+l_2-L)}A_{l_1l_2}^{\ell M}, \\ \nonumber
A_{ll}^{\ell M} \, &=& \, A_{ll}^{\ell M} \,\,\delta_{\ell,2k}, \,\,\,\,\,\,\,\,\,\,\,\,\,\,\,\,\,\,\,\,\, k=0,\,1,\,2,\,\cdots.
\end{eqnarray} 

The Bipolar Spherical Harmonic (BipoSH) coefficients,
$A_{l_1l_2}^{\ell M}$, are linear combinations of off-diagonal
elements of the  harmonic space covariance matrix,
\begin{equation}\label{ALMvsalm}
A^{\ell M}_{l_1 l_2}\,=\, \sum_{m_1m_2} \langle a_{l_1m_1}a^{*}_{l_2 m_2}\rangle (-1)^{m_2} C^{\ell M}_{l_1m_1l_2 -m_2}.
\end{equation}
This means that $A^{\ell M}_{l_1 l_2}$ completely represent the
information of the covariance matrix in harmonic space $\langle
a_{l_1m_1}a^{*}_{l_2 m_2}\rangle$.  When SI holds, the harmonic space
covariance matrix is diagonal and hence
\begin{eqnarray}  \label{SIALM}
A_{ll^\prime}^{\ell M}\,&=&\,(-1)^l C_{l} (2l+1)^{1/2} \,  \delta_{ll^\prime}\, \delta_{\ell 0}\, \delta_{M0},
\\ \nonumber
A^{0 0}_{l_1 l_2}\,&=&\, (-1)^{l_1} \sqrt{2l_1+1}\, C_{l_1}\, \delta_{l_1l_2}.
\end{eqnarray}  

BipoSH expansion is the most general representation of the two point
correlation functions of CMB anisotropy. The well known angular power
spectrum, $C_l$ is a subspace of BipoSH coefficients corresponding to
the $A_{ll}^{00}$ that represent the statistically isotropic part of a
general correlation function. When SI holds, $A_{ll}^{00}$ or
equivalently $C_l$ have all the information of the field. But when SI
breaks down, $A_{ll}^{00}$ are not adequate for describing the field,
and one needs to take the other terms into account.  The Bipolar power
spectrum (BiPS) is defined as a rotationally invariant contraction of
the BipoSH coefficients
\begin{equation}\label{kappal}
\kappa_\ell \,=\, \sum_{l,l',M} |A_{ll'}^{\ell M}|^2 \geq 0.
\end{equation} 
This definition is identical to the real space expression in
eq.~(\ref{kl}). More importantly, BiPS is measurable from a single CMB
map since averages over many independent modes and reduces the cosmic
variance~\footnote{This is similar to combining $a_{lm}$ to construct
the angular power spectrum, $C_l=\frac{1}{2l+1}\sum_{m}{|a_{lm}|^2}$,
to reduce the cosmic variance}~\cite{us_apjl}. The BiPS of the CMB
anisotropy maps measured by WMAP has been computed\cite{us_apjl2}.
Preliminary BiPS results on the CMB polarization maps from the three
year of WMAP data have also emerged in past few
months~\cite{bas06,uci_si06}.

The BiPS is sensitive to structures and patterns in the underlying
total two-point correlation function \cite{us_apjl, us_pascos}.  The
BiPS is particularly sensitive to real space correlation patterns
(preferred directions, etc.) on characteristic angular scales. In
harmonic space, the BiPS at multipole $\ell$ sums power in
off-diagonal elements of the covariance matrix, $\langle a_{lm}
a_{l'm'}\rangle$, in the same way that the `angular momentum' addition
of states $l m$, $l' m'$ have non-zero overlap with a state with
angular momentum $|l-l'|<\ell<l+l'$. Signatures, like $a_{lm}$ and
$a_{l+n m}$ being correlated over a significant range $l$ are ideal
targets for BiPS. These are typical of SI violation due to cosmic
topology and the predicted BiPS in these models have a strong spectral
signature in the bipolar multipole $\ell$ space~\cite{us_prl}.  The
orientation independence of BiPS is an advantage since one can obtain
constraints on cosmic topology that do not depend on the unknown (but
specific) orientation of the pattern ({\it{e.g.}}, preferred
directions of DD relative to the sky).

Spaces of constant curvature have been completely
classified~\cite{wol94vin93,thur79}.  For Euclidean geometry, there
are known to be six possible topologies that lead to orientable
spaces.  The simple flat torus, ${\cal M} = T^3$, is obtained by
identifying the universal cover ${\cal M}^u={\cal E}^3$ under a
discrete group of translations along three non-degenerate axes,
${\mathbf s_1,\mathbf s_2,\mathbf s_3}$: $ {\mathbf s_i} \to {\mathbf
s_i} + {\mathbf n} L_i$, where $L_i$ is the identification length of
the torus along $s_i$ and ${\bf n}$ is a vector with integer
components.  In the most general form, the fundamental domain (FD) is
a parallelepiped defined by three sides $L_i$ and the three angles
$\alpha_i$ between the axes (`squeezed torus').  If ${\mathbf s_i}$ are
orthogonal then one gets cuboid FD, which for equal $L_i$ reduces to
the cubic torus.  The cuboid and squeezed spaces which can be obtained
by a linear coordinate transformation ${\cal L}$ on cubic torus can
have distinctly different global symmetry~\footnote{For cubic torus
the Dirichlet domain (DD) matches the fundamental domain (FD).
However, for torus spaces with cuboid and parallelepiped FD, the
corresponding DD is very different, e.g., hexagonal
prism~\protect{\cite{wol94vin93,us_prl}}.}.

\begin{figure}[h]
  \begin{center}
\includegraphics[scale=0.5,
angle=0]{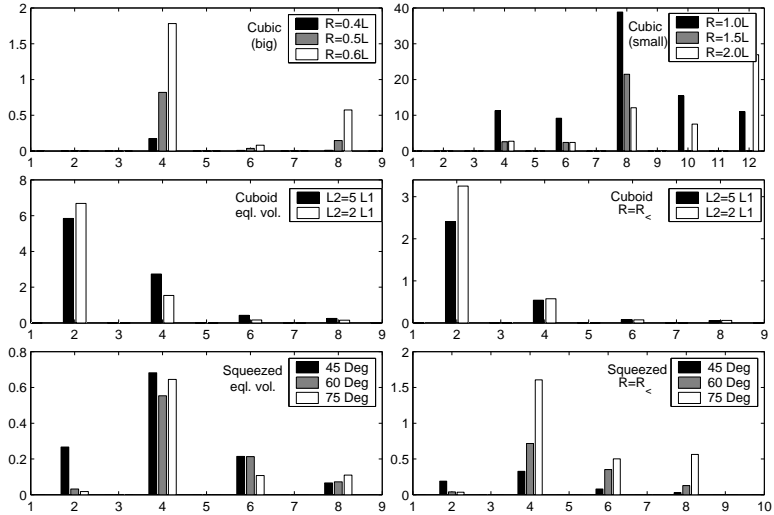}  
  \end{center}
  \caption{\footnotesize The figure taken from~\protect\cite{us_prl}
    shows the BiPS $\kappa_{\ell}$ spectra for flat tori models .  The
    top row panels are for cubic tori spaces. The left panel shows
    spaces of volume, $V_{\cal M}$, larger than the volume $V_{*}$
    contained in the sphere of last scattering (SLS) with $V_{\cal
    M}/V_{*} = 3.7,1.9,1.1$, respectively.  The right panel shows
    small spaces with $V_{\cal M}/V_{*} = 0.24,0.07,0.03$,
    respectively.  Note that $\kappa_2=0$ for cubic tori.  The middle
    panels consider cuboid tori with $1:5$ and $1:2$ ratio of
    identification lengths. The bottom panels show $\kappa_{\ell}$ for
    equal-sided squeezed tori with $\alpha=45^\circ,60^\circ$ and
    $75^\circ$. In the middle and bottom rows, the right panels show
    the case when radius of SLS, $R_*=R_<$ the in-radius of the space.
    Here, the SLS just touches its nearest images which is at the
    threshold where CMB anisotropy is multiply imaged for larger
    $R_*$. The cases in the left panels of lower two rows have
    $V_{\cal M}/V_{*} =1$ and are at the divide between large and
    small spaces. }
  \label{kappaltorus}
\end{figure}

 Study of the BiPS signature of cosmic topology has already been
undertaken and ongoing~\cite{us_prl,us_big,us_dodeca}.  The
correlation function $C({\hat q,\hat q^\prime})$ for CMB anisotropy in
a multiply-connected universe such as the torus space can be computed
using the regularized method of images\cite{bps00a}.
Fig.~\ref{kappaltorus} plots the predicted $\kappa_\ell$ spectrum for
a number of cubic, cuboid and squeezed torus spaces~\cite{us_prl}.
Similar results for Poincar\'e Dodecahedron show a characteristic BiPS
with a dominant peak at $\kappa_6$ over a range of value of curvature
radius (including integrated Sachs-Wolfe effect). This relates to the
angular separation of the directions to faces of the extremely
symmetric DD of Poincar\'e dodecahedral space.  Fig.~\ref{kappaltorus}
shows that $\kappa_\ell$ is zero for odd $\ell$.  This is intimately
related to the symmetries of the Dirichlet domain, which in turn is
dictated by the properties of the subgroup of isometries $\Gamma$.
The BiPS is zero at odd bipolar multipoles if DD has $2n$-fold
symmetry about an axis and reflection symmetry in an orthogonal plane.
It can be proved that all Euclidean and all spherical spaces generated
by single action $\Gamma$ satisfy this condition. Remarkably enough,
compact hyperbolic spaces do not satisfy these condition, and are
generically expected to have non-zero BiPS at odd value bipolar
multipoles $\ell$~\cite{him06}.  This provides a {\em measurable}
classification of cosmic topology based on CMB anisotropy and
polarization, {\em i.e., a spectroscopy of cosmic topology.}

A simple working example is the BiPS signature of a non-trivial
topology can be given for a $T^3$ universe, where the correlation
function is given by
 \begin{equation}
C({\hat q,\hat q^\prime}) = L^{-3} \sum _{{\bf n}}
P_\Phi(k_{\bf n}) \,\,{\mathrm e}^{-i \pi 
(\epsilon_{\hat q} {\bf n}\cdot {\hat q} - \epsilon_{\hat q^\prime} 
{\bf n}\cdot 
{\hat q}^\prime)},
\label{C_tor}
\end{equation}
in which, ${\bf n}$ is 3-tuple of integers (in order to avoid
confusion, we use $\hat{q}$ to represent the direction instead of
$\hat{n}$), the small parameter $\epsilon_{\hat q} \le 1 $ is the
physical distance to the SLS along $\hat q$ in units of $L/2$ (more
generally, $\bar L/2$ where $\bar L= (L_1L_2L_3)^{1/3}$) and $L$ is
the size of the Dirichlet domain (DD). When $\epsilon$ is a small
constant, the leading order terms in the correlation function
eq.~(\ref{C_tor}) can be readily obtained in power series expansion in
powers of $\epsilon$.  For the lowest wave numbers $|{\mathbf n}|^2=1$
in a cuboid torus~\cite{us_prl}
\begin{eqnarray} 
C({\hat q,\hat q^\prime}) &\approx& 2 \sum_i P_\Phi({2\pi}/{L_i}) 
\cos(\pi\epsilon\beta_i\Delta q_i) \\ \nonumber
&\approx& C_0 \left[1 - \epsilon^2\, 
|{\mathbf \Delta q}|^2 + 3\,\epsilon^4 \, \sum_{i=1}^3
(\Delta q_i)^4   \right],
\label{appcorr}
\end{eqnarray} 
where $\Delta q_i$ are the components of ${\mathbf \Delta q} =\hat
q-\hat q^\prime$ along the three axes of the torus and $\beta_i = \bar
L /L_i$. From this, the non-zero $\kappa_\ell$ can be analytically
computed to be
\begin{eqnarray}
\frac{\kappa_0}{C_0^2}\, &=&\,\pi^2(1-4\epsilon^2 
+\frac{368}{15} \epsilon^4-\frac{288}{5}\epsilon^6+\frac{20736}{125}\epsilon^8)
\nonumber \\
\frac{\kappa_4}{C_0^2}\, &=&\, \frac{12288 \pi^2 }{875} \epsilon^8\,\,.  
\end{eqnarray}
$\kappa_4$ has the information of the relative size of the Dirichlet
domain and one can use it to constrain the topology of the universe.

The results of WMAP are a milestone in CMB anisotropy measurements
since it combines high angular resolution, high sensitivity, with
`full' sky coverage allowed by a space mission.  The {\it Wilkinson
Microwave Anisotropy Probe} ({\it WMAP}) observations are consistent
with the predictions of the concordance $\Lambda$CDM model with
scale-invariant and adiabatic fluctuations which have been generated
during the inflationary epoch \cite{ben_wmap03,hin_wmap06,
sper_wmap06}. After the first year of {\it WMAP} data, the SI of the
CMB anisotropy ({\it i.e.} rotational invariance of n-point
correlations) has attracted considerable attention.  Tantalizing
evidence of SI breakdown (albeit, in very different guises) that
mounted in the {\it WMAP} first year sky maps, using a variety of
different statistics are expected to persist in the three persist in
three year data~(see \cite{uci_si06} for discussion and references.).

The CMB anisotropy map based on the WMAP data are ideal for testing
for statistical isotropy. Preferred directions and statistically
anisotropic CMB anisotropy have been discussed in literature earlier
\cite{fer_mag97,bun_scot00}. A number of direct searches for signature
of cosmic topology have been proposed and carried out on early CMB
data from COBE-DMR. Full Bayesian likelihood comparison to the data of
specific cosmic topology models is another approach that has applied
to COBE-DMR data~\cite{bps98,bps00a,bps00b}.  The generic features of
$\kappa_\ell$ spectrum are related to the symmetries of correlation
pattern. For cosmic topology, $\kappa_\ell$ are sensitive to SI
violation even when CMB is not multiply imaged.  The orientation
independence of BiPS is an advantage for constraining patterns
(preferred directions) with unspecified orientation in the CMB sky
such as that arising due to cosmic topology or, anisotropic
cosmology~\cite{ghos06}. Extension of BiPS analysis to CMB
polarization maps has been studied recently~\cite{bas06,uci_si06} adds
a new dimension to the spectroscopy of cosmic topology.

In summary, there are strong theoretical and philosophical motivations
for a non-trivial cosmic topology.  The breakdown of statistical
homogeneity and isotropy of cosmic perturbations is a generic feature
of non trivial cosmic topology.  A promising observational approach is
to hunt for SI violation in the CMB anisotropy.  The underlying
correlation patterns in the CMB anisotropy and polarization in a
multiply connected universe is related to the symmetry of the
Dirichlet domain.  BiPS has the advantage of being independent of the
overall orientation of the Dirichlet domain with respect to the
sky. The pattern of SI violation of a {\em cosmic topology leads to a
measurable, characteristic Bipolar power spectrum} related to the
principle directions in the Dirichlet domain and symmetries of the two
point correlation function. The Bipolar power spectroscopy of cosmic
topology presents itself as promising pursuit for current and upcoming
measurements of CMB anisotropy and polarization.

The author acknowledges years of very fruitful collaboration with Dick
Bond and Dmitry Pogosyan.  Amir Hajian devoted most of his doctoral
thesis to this effort and the results mentioned here were jointly
obtained with him The article also includes recent and ongoing work
done with Himan Mukhopadhyay and Soumen Basak.


\begin{thebibliography}{99}

\bibitem
{ell71} G. F. R. Ellis,  Gen. Rel. Grav. {\bf 2}, 7 (1971).

\bibitem
{sok_shv74} D. D. Sokolov and
V. F. Shvartsman (1974) Zh. Eksp. Theor. Fiz. {\bf 66}, 412  [JETP,
{\bf 39}, 196 (1974)].

\bibitem
{got80} J. R. Gott, Mon. Not. R. Astr. Soc. {\bf 193}, 153 (1980).

\bibitem
{lac_lum95} M. Lachieze-Rey and J. -P.  Luminet, Phys. Rep.  {\bf 25},
136, (1995).

\bibitem
{stark98}G.  Starkman,  Class., Quantum
  Grav. {\bf 15}, 2529 (1998).

\bibitem
{linde} A. Linde JCAP 0410, 004, (2004)

\bibitem
{lev02} J. Levin, Phys. Rep. {\bf 365}, 251, (2002).

\bibitem
{sour00} T.~Souradeep, in `The Universe', eds. Dadhich, N.  \&
  Kembhavi, A., (Kluwer 2000).
\bibitem{bps98} J. R. Bond, D. Pogosyan \& T. Souradeep, Class. Quant.
  Grav. {\bf 15}, 2671 (1998).

\bibitem
{staro} A.  de Oliveira-Costa, G. F. Smoot, A. A. Starobinsky, , ApJ
{\bf 468}, 457 (1996)

\bibitem
{levin98} J. Levin, E. Scannapieco, J. Silk , Class.Quant.Grav. {\bf
15}, 2689, (1998).

\bibitem{bps00a} J. R. Bond, D. Pogosyan \& T. Souradeep, Phys.  Rev. {\bf D
  62},043005 (2000).

\bibitem{bps00b} J. R. Bond, D. Pogosyan \& T. Souradeep, Phys.
Rev. {\bf D 62},043006 (2000).

\bibitem
{angelwmap} A.~de Oliveira-Costa,
M.~Tegmark, M.~Zaldarriaga, \& A.~Hamilton, 2004, Phys. Rev.{\bf D69}, 063516.

\bibitem
{coles-graca} P. Dineen, G. Rocha, P. Coles,
Mon.Not.Roy.Astron.Soc. {\bf 358}, 1285 (2005)

\bibitem
{Copi:2003kt}
  C.~J.~Copi, D.~Huterer and G.~D.~Starkman,
  Phys.\ Rev.\ D {\bf 70}, 043515 (2004)

\bibitem
{pag06} L. Page et al., {\it preprint} 2006.


\bibitem
{us_prl} A. Hajian, \& T. Souradeep, {\it preprint}
(astro-ph/0301590).

\bibitem
{us_apjl} A. Hajian and T. Souradeep ApJ Lett. 597, L5 (2003).

\bibitem
{Var} D. A. Varshalovich, A. N.Moskalev, V. K.Khersonskii, {\it
  Quantum Theory of Angular Momentum} (World Scientific, 1988).


\bibitem{us_apjl2} A. Hajian \&  T. Souradeep \& N. Cornish,
  Astrophys. J. Lett. {\bf 618}, L63 (2005). 

\bibitem{bas06}
S. Basak, A. Hajian \&  T. Souradeep, Phys. Rev. {\bf D 74},
021301(R) (2006).


\bibitem{uci_si06} T. Souradeep, A. Hajian and S. Basak, {\it preprint}
  Proc. of `Fundamental Physics With CMB workshop', UC Irvine, March
  23-25, 2006, to be published in New Astronomy Reviews
  (astro-ph/0607577).


\bibitem
{us_pascos}T. Souradeep, and
A. Hajian  Pramana, {\bf 62}, 793, (2004).

\bibitem
{wol94vin93} J. A. Wolf, {\it Space
    of Constant Curvature (5th ed.)}, (Publish or Perish, Inc., 1994);
    E. B. Vinberg, {\it Geometry II -- Spaces of constant curvature},

\bibitem
{thur79}W. P. Thurston, {\it The Geometry of 
3-Manifolds},lecture notes, (Princeton University 1979).
\bibitem
{us_big} A.  Hajian, \& T. Souradeep, {\it preprint}
(astro-ph/0501001)

\bibitem
{us_dodeca} A.Hajian, D.  Pogosyan, T. Souradeep, C.Contaldi,
J. R. Bond, {\it in preparation}; Proc. 20th IAP Colloquium on Cosmic
Microwave Background physics and observation, 2004.

\bibitem
{him06} H. Mukhopadhyay and  T. Souradeep, {\it in preparation}.
\bibitem{ben_wmap03} C. L. Bennett et al., Astrophys.J.Suppl. {\bf
148}, 1, (2003).

\bibitem{hin_wmap06} G. Hinshaw  et al., {\it preprint} ,  (astro-ph/0603451).
\bibitem{sper_wmap06} D. Spergel et al., {\it preprint}, (astro-ph/0603449).


\bibitem
{fer_mag97} P. G. Ferreira \&
  J. Magueijo, Phys.  Rev.  {\bf D56}, 4578, (1997).

\bibitem
{bun_scot00} E. Bunn \& D. Scott,
  M.N.R.A.S., {\bf 313}, 331, (2000).

\bibitem[{Ghosh et al.} (2006)]{ghos06} T.~Ghosh, A.~Hajian and
  T.~Souradeep,{\it preprint} [arXiv:astro-ph/0604279].

\end{thebibliography}
\end{document}